\documentclass[10pt,aps,prl,amssymb,amsmath,amsfonts,twocolumn,superscriptaddress,showpacs]{revtex4-1}
\usepackage{graphicx}

\begin{document}

\title{Fluctuation-induced traffic congestion in heterogeneous networks}

\author{A. S. Stepanenko}
\affiliation{School of Engineering and Applied Science, Aston University, Birmingham B4 7ET, UK}
\author{I. V. Yurkevich}
\affiliation{School of Physics and Astronomy, University of Birmingham,
Birmingham B15 2TT, UK}
\author{C. C. Constantinou}
\affiliation{School of Electronic, Electrical \& Computer Engineering, University of Birmingham, Birmingham B15 2TT, UK.}
\author{I. V. Lerner}
\affiliation{School of Physics and Astronomy, University of Birmingham,
Birmingham B15 2TT, UK}

\begin{abstract}
    \noindent In studies of complex heterogeneous networks, particularly of the Internet, significant attention was paid to analyzing network failures caused by hardware faults or overload, where the network reaction was modeled as rerouting of traffic away from failed or congested elements. Here we model another type of the network reaction to congestion -- a sharp reduction of the input traffic rate through congested routes which occurs on much shorter time scales.    We consider the onset of congestion in the Internet where local mismatch between demand and capacity results in traffic losses and show that it can be described as a phase transition characterized by strong non-Gaussian loss fluctuations at a mesoscopic time scale.  The fluctuations, caused by noise in input traffic, are exacerbated by the heterogeneous nature of the network manifested in a scale-free load distribution.  They result in the network strongly overreacting to the first signs of congestion by significantly reducing input traffic along the communication paths where congestion is utterly negligible.
\end{abstract}


\maketitle

All Internet users are familiar with the feeling of frustration when their network connection slows down or halts. Barring cascading  failures \cite
{Watts:02,BBV:08,Motter:02,*Motter:03,Moreno:03,Ashton:05,WangKim:07,Havlin:2010,*Vespignani:10}  which can shut down parts of the network, such a slowdown is a
sign of network congestion which  happens when the traffic load on some network elements exceeds their capacity  \cite{BBV:08,Motter:02,*Motter:03,Moreno:03,WangKim:07}.
For the Internet  congestion can be quantified    as a relative difference between the rates of sent and delivered data packets \cite{BBV:08,Arenas:01}, with excess  packets being eventually dropped.
The first network reaction to a lost packet is a significant reduction of a transmission rate at the source followed by a slow recovery to the normal rate. When several loss events occur in quick succession,  a multiplicative reduction drastically suppresses the transmission rate, which feels as congestion for the end user.  If congestion persists for longer, the network eventually reroutes traffic away from congested links which may overload other links triggering a cascade of failures  \cite{Moreno:03}.

A surprising result of the considerations presented here is that transmission rates might be  significantly reduced  when the relative number of lost packets is utterly negligible.
Such a reduction results from the existence of strong fluctuations of data losses along a typical communication path at the very onset of congestion. The loss fluctuations arise at each link at the threshold of its capacity due to noise in input traffic.  Although such fluctuations exist only on shorter, \emph{mesoscopic}  time scales   and will die out in time, we show that they might trigger an overreaction of a typical  transport protocol   to the first signs of losses. Normally the  protocol aggressively reduces traffic rates along the routes perceived as congested due to multiple loss events. The overreaction results from the probability of such events on the mesoscopic scale being much higher than the product of the single-event probabilities.

\begin{figure}[b]
\begin{center}
\includegraphics[width=0.9\columnwidth]{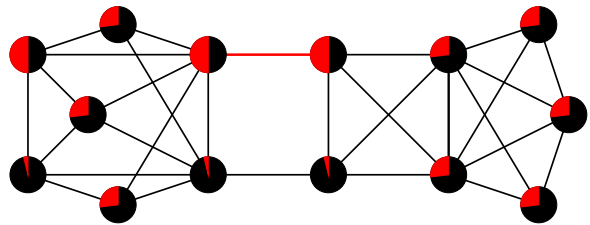}
\end{center}
\caption{The importance of links with high betweenness (load): one congested link ($3\%$ of all the links here) affects $27\%$ of all the shortest paths. A light-shaded sector on each node indicates the fraction  of congested paths originated on it. }
\label{Fig1}
\end{figure}

The fluctuations are greatly magnified in heterogenous networks characterized by a power law (PL) distribution of the link load  since congestion on links with high load affects a disproportionately large number of communication paths, as illustrated in
Fig.~\ref{Fig1}. The link  load in a network with a homogeneous traffic input distribution is proportional to the \emph{link betweenness} $B_i$ (roughly, the number of shortest paths through  link $i$)   \cite{Freeman:77,*Newman:01,*Newman:02}.   Many heterogeneous networks, including the Internet, fall into the category of scale-free (SF) networks characterized by the PL distribution of node degree \cite{Barabasi:99,Faloutsos:1999,Vespignani:01,Albert:02,P-SV:04}.
Load   distribution in SF networks also follows a (truncated) PL \cite{P-SV:04,Goh:01,*Goh:02,*Barthelemy:03,*Goh:03,Boccaletti:06,*Dorogovtsev:08},
\begin{align}\label{PB}
    {\mathcal{P}}_{\mathrm{L}}(\ell )&\propto  \ell ^{-2-\delta} \,,
\end{align}
with an almost universal exponent, $2+\delta \sim 2.0 \div2.3$.  The load distribution of the real Internet was found \cite{P-SV:04,Vazquez:02,Dimitropoulos:05} to be in agreement with the above.

We focus  on a \emph{critical regime} at the  onset of congestion with a small
imbalance, $\eta_i\equiv1-\tau_ir_i$, between the average packet  arrival rate (load), $1/\tau_i$, and  departure rate (capacity), $r_i$, at link $i$.  The nodes  in the Internet core  are routers and the links are output memory buffers (with attached transmission lines). For  $\eta_i>0$  a memory buffer eventually becomes full and a newly arrived packet is dropped.  On average, $\eta_i\!=\!\left\langle \Phi_i(T) \right\rangle$ where $\Phi_i(T)$ is the fraction of packets dropped  during an observation time window $T$. Shifting this  window in time causes $\Phi$ to fluctuate due to inevitable flow fluctuations \cite{Menezes,*Meloni2008}. We show the loss fluctuations to be crucial for network transport at certain \emph{mesoscopic} time scale; for large enough $T$ they  die out  and $\Phi_i$   equals    $\eta_i$.
Positive $\eta_i$ plays the role of a local congestion parameter: their sum defines the network congestion parameter \cite{BBV:08,Arenas:01}.

Relative loss, $\Phi $,  along a typical  communication path is governed by losses  in the comprising links and fluctuates due to both the noise in each link and a random choice of links in the path. The probability of a  randomly picked link to be in the path is proportional to its betweenness. Hence, in a network with average link betweenness $\bar B$, a \emph{small} loss along a path with $a$ links is given by
\begin{align}\label{typ}
{\Phi} &=  \sum_{i=1}^a {\ell }_i   \Phi_i \,,&  {\ell }_i &\equiv  {B _i}/{ \bar B    }\,,
\end{align}
The quenched distribution of the relative load $\ell_i$ is   given by the truncated PL (\ref{PB}), cut from below by $\ell\!\sim\!1$.  The upper cutoff is irrelevant for $\delta>0$.

To describe noise in packet arrivals at link $i$ we assume, without loss of generality, that  the inter-arrival time is random, with average $\tau_i $, while  packets have a fixed length $l_0$. Arriving packets join  the  queue in the memory buffer. The queue length, $x_i({t})$ (measured in $l_0$), performs a random walk bounded  by a buffer size  $c_i$. The  probability density, $w_i$,  of diffusion from $x'$ to $x$ over time $t$ obeys the Fokker-Planck equation with diffusion and advection coefficients $D_i \equiv1/\tau_i  $ and $V_i\equiv \eta_i/\tau_i $ and the probability-conservation boundary conditions:
\begin{equation}
 \partial_t w_i (x, x'; t)
  =\left [{- V _i\partial_{x }
    + D_i \partial_{x }^2}\right]w_i(x, x'; t)\, .
\label{FP}
\end{equation}

In the critical regime  the queue hovers at the boundary. A newly arriving packet is dropped every time when the queue length $x_i(t)$ overflows reaching the boundary layer, $c_i-1\leqslant x_i(t)\leqslant c_i$. Thus the fraction of  packets lost over an observation time $T\!\gg\!\tau_i$ is
\begin{align}\label{Phin}
   \Phi_i({T}) &= \frac{1}{T} \int _{0}^{T}{\mathrm{d}}t\, \theta\left[x_i({t})-c_i+1\right] =\frac{1}{K_i}\sum_{k=1}^{K_i}\delta_k \,,
\end{align}
where $ K_i  \equiv \left[{T }/ {\tau_i}\right]\gg1$ is the number of packets arrived over time $T$, and $\delta_k $ equals $1$ if $x_i({t})$ reached the boundary layer at the $k^{\mathrm{th}} $ step,  or  $ 0$ otherwise.

To find the probability density function (PDF) of $\Phi_i $, we begin with the  characteristic function,
$
\chi_T (q_i)=\langle\mathrm{ e}^{\mathrm{i}q_i\Lambda_i  }\rangle\,
$, of the distribution of a cumulative loss,  $\Lambda_i \!=\!   (T/\tau_i)\Phi_i(T) $. Using Eqs.~(\ref{FP}) and ({\ref{Phin}})   we represent  $\chi _T$  as the sum of time-ordered integrals
\begin{align*}
   \chi_T(q_i)=\sum_{n=0}^ \infty({{\mathrm{i}q_i} })^n \! \!\int\!\mathrm{d}^n t\, {\mathcal{R}}_i({t_n\!-\!t_{n-1}})\dots  {\mathcal{R}}_i({t_2\!-\!t_1}) p_i\,,
\end{align*}
running over the regions $0<t_1<\dots < t_n<T$. Here $p_i\equiv w_i({c_i,x' ; t\to\infty})= \eta_i \left(1-\mathrm{e}^{-\eta_ic_i}\right)^{-1}$ is  the stationary probability density for the queue to be in the boundary layer and  ${\mathcal{R}}_i(t)=w({c_i,c_i;t})$  is the return probability.  We rewrite the expression for
 $\chi_T $ as the integral  equation
\begin{align}\label{chi}
{\chi}_T(q_i)\!-\!1={\mathrm{i}} q_i\biggl\{p_i T+\!\!\int_{0}^{T}\!\! \mathrm{d} t\,{\mathcal{R}_i}(T\!-\!t)\,\left[{\chi}_{t}(q_i)-1\right] \!\biggr\}.
\end{align}
The inverse Fourier transform from $q_i$ to $\Lambda_i$ shows  the PDF of $\Lambda_i$ to be the sum   $\mathcal{F}_T({\Lambda_i })+A _i\delta({\Lambda_i})$, with $\mathcal{F} $  describing losses ($\Lambda_i>0$) and  $A_i$  the probability of   no losses  over the time $T$, with
  $1=A_i+\int_0^\infty\!\mathrm{d}\Lambda\ {\mathcal  F}_T (\Lambda)$. Solving Eq.~(\ref{chi})  by  the Laplace transform  with respect to $T$  gives
\begin{align}\label{F}
{\mathcal  F}_{\varepsilon}(\Lambda _i )&=\frac{  p_i \mathrm{e}^{-\frac{\Lambda _i }{\mathcal{R}_{\,\varepsilon}}}}{\tau _i \varepsilon^2\,\mathcal{R}_{\,\varepsilon}^2}\,
,&
\mathcal{R}_{\,\varepsilon}&=\frac{   \sqrt{  \eta_i^2+4 \tau _i {\varepsilon}  }  + {  \eta_i}}{2\tau_i  \varepsilon}  \,.
\end{align}

For a \emph{perfectly designed} network with fully utilized resources  for homogeneous input traffic,  $c_i$ and  $ \tau_i^{-1}$ are proportional to the relative link load \cite{Goh:01},   $c_i\equiv c {\ell}  _i $   and $ \tau_i^{-1} \equiv {\ell} _i  \tau^{-1}$, while  the imbalance $\eta_i$ is   $\ell _i$-independent.    Then the congestion threshold  $\eta_i=0$ is reached simultaneously by all  links. Naturally, such a complete utilization is impossible:  design imperfections and local variations of demand cause the congestion thresholds to spread \cite{WangKim:07}.
We  model such a spread (quenched on the relevant time scales) as a sharply peaked symmetric distribution of $\eta_i$ with  \emph{criticality width} $\gamma\ll1$.  Realistic  regimes are bounded by $a\ll \gamma^{-1} \lesssim c  $, as for $\gamma^{-1} \gtrsim c$  congestion thresholds are still reached  simultaneously, while a network with $a\gamma\gtrsim1 $ would be permanently congested.

The  PDF of $\Phi _i$ has the same form as that of $\Lambda _i$, namely   $A_i\delta({\Phi_i})+{P}_T(\Phi_i;\ell_i) $, where its lossy part,  ${P}_T(\Phi_i;\ell_i)$,  is  found for fixed $\ell_i$ and $\eta_i$ by rescaling    $\mathcal{F}_T $, the inverse Laplace transform of Eq.~(\ref{F}), as
$
 (\ell _i/\varphi_0^2){\mathcal F}_T \! \left( \ell _i\Phi _i/\varphi_0^2\right)
$.  The averaging over $\eta $ is straightforward and preserves very different shapes of $P_T$ on the \emph{mesoscopic time scale}, $\ell_i\gamma^2\!\ll\! \tau/T\!\equiv\!\varphi_0^2\!\ll\! \gamma$, and \emph{macroscopic} one, $\varphi_0^2\!\ll\! \gamma$. In the former case, the congestion spread is so narrow  that to average over it  one replaces    $\eta_i$ by $\gamma$ which gives  the averaged PDF
  $ \sim\ell _i\gamma /\varphi_0^2$ for $0<\Phi _i \lesssim \varphi_0/\sqrt{ \ell _i}$ followed by a decay  $\sim \mathrm{e}^{-\Phi _i^2  \ell _i/\varphi_0^2} $ for bigger $\Phi_i$. The probability of loosing a packet, $1-A_i\sim\gamma\sqrt{ \ell_i }/\varphi_0\ll1$, is small.
For macroscopic times, the Laplace transform in Eq.~(\ref{F}) is  exponential, corresponding to the averaged PDF given by $\delta({\Phi_i-\eta_i})$ for $\eta_i>0 $ \footnote{More accurate calculations show that the $\delta$-function is smeared but the resulting width is much smaller than $\gamma$.}: on this scale ${P}_T(\Phi_i;\ell_i) $ repeats the shape of the criticality spread for positive $\eta_i$,   while free-flow (lossless) links with negative $\eta_i$   give $A_i=\frac{1}{2}$.

\begin{figure}[t]
\begin{center}
\includegraphics[width=0.8\columnwidth]{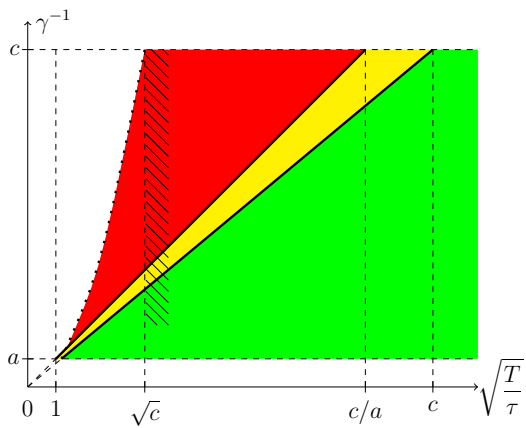}
\end{center}
\caption{ Different loss modes depending on the operation time $T$ and  {criticality width} $\gamma$.    The dark-shaded (red online) area above the upper bold line, $\gamma^{-1}=a\sqrt{T/\tau} $, represents the fluctuation-driven mesoscopic mode;  the light-shaded (green online) area below the lower bold line, $\gamma^{-1}= \sqrt{T/\tau} $,  the self-averaging macroscopic mode (microscopic times on the left of the parabola,  $\gamma^{-1}=T/\tau$, are not considered). The crossover sector between the bold lines is  narrow as $c\gg a\gg1$ ($a$ is the number of links in path ({\ref{typ}}), $c$ is the memory buffer size for link with $\ell _i=1$).   The hatched area shows the region  of network feedback operations at the congestion onset. }
\label{Fig2}
\end{figure}

\begin{figure}[t]
\includegraphics[width=.9\columnwidth]{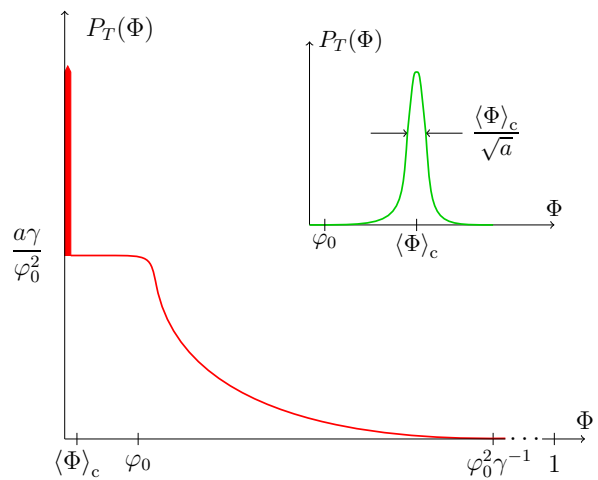}
\caption{Probability distribution function, $P_T({\Phi})$, of relative losses in the critical regime at the onset of congestion. The average, $\left\langle \Phi \right\rangle_\mathrm{c}\sim a\gamma \ll1$, is time-independent. The main figure shows (not to scale)  $P_T({\Phi})$ when the observation time $T$ is on the mesoscopic scale. With increasing $T$, as $\varphi_0\!\equiv\! \sqrt{\tau/T}$ is moving towards $\left\langle \Phi \right\rangle_\mathrm{c}$, the plateau becomes narrower and higher, the tail squeezes, and the area of the peak at $\Phi=0$ shrinks. After passing through the intermediate time scale, $\langle{\varphi_0}\rangle$  becomes larger than $\left\langle \Phi \right\rangle_\mathrm{c}$ on the macroscopic time scale,  where the PDF becomes Gaussian as shown in the inset.
}\label{Fig3}
\end{figure}

The PDF of losses  along   path (\ref{typ}) is a convolution of PDFs of statistically independent ${\ell} _i\Phi _i$, averaged over the load distribution (\ref{PB}). It still has the  structure  $A\delta({\Phi})+P_T({\Phi})$   with $A=\prod_{i=1}^a\left\langle A_i \right\rangle_\ell$.
For  $\varphi_0 \ll\gamma$ losses in each link in  (\ref{typ}) are on the  \emph{macroscopic} time scale, light-shaded (green online) in Fig.~\ref{Fig2}. Then $A =2^{-a}\ll1$ for $a\gg1$ and, by the law of large numbers, $P_T({\Phi})$ is a normal distribution  with average $ \left\langle\Phi \right\rangle_\mathrm{c}\sim a\gamma\ll1$ (where $\left\langle \dots \right\rangle_\mathrm{c}$ stands for the averaging over all the three sources of randomness) and width $\sim \left\langle\Phi \right\rangle_\mathrm{c} /\sqrt{a}$  (inset in Fig.~\ref{Fig3}).

The communication path  is in a nontrivial \emph{fluctuation-driven} mode only if all the comprising links are in that mode, obeying $\ell_i\ll \varphi_0^2/\gamma^2$. For $\delta>0$ in Eq.~(\ref{PB}) and $a\gamma\!\ll\!\varphi_0 \!\ll\!\sqrt{\gamma} $, which defines the \emph{mesoscopic time scale} for the  path  shown as the dark-shaded area (red online) in Fig.~\ref{Fig2},  one has $A=\left({1-\gamma/\varphi_0}\right)^a\approx1-a\gamma/\varphi_0$.
The PDF has  a  peak at $\Phi \!=\!0$, describing the close-to-$1$ probability $A$ of not  loosing  a  packet, while the lossy part,  $P_T({\Phi})$, is  dominated   by (any) one link along the path: $P_T({\Phi }) \cong a  \left\langle { {\ell }^{-1} P_T   ( \Phi/{\ell } ; \ell  )   } \right\rangle_\ell  $.
The plateau in this rescaled PDF is  stretched  up to $ \varphi_0\ell ^{1/2} $ with height $a\gamma/\varphi_0^2$. For $\Phi \lesssim\varphi_0$  averaging over the  distribution  (\ref{PB}) leaves the plateau intact. For   $\Phi \gtrsim \varphi_0$ the plateau exists only for links with   load $ {\ell }\gtrsim ({\Phi /\varphi_0})^2$ which becomes the lower limit in the averaging over load.  When the lower limit is much smaller than the upper, $ ({\varphi_0/\gamma})^2$, the averaging is contributed only by the former resulting in  the  PL tail
\begin{align}\label{tail}
   P_T(\Phi)&\sim  \frac {a\gamma}{\varphi_0^2}
   \left(\frac{\Phi}{\varphi_0}\right)^{-2(1+\delta)}\!\!\!,
   &   {\varphi_0}\lesssim  {\Phi } \ll \frac {\varphi_0^2}{\gamma}\,,
\end{align}
followed by an exponentially small decay.

The entire PDF is shown in Fig.~\ref{Fig3}. The probability of losses is small but when losses do occur -- the conditional probability of higher-than-average losses (governed by the long plateau and even longer heavy tail)  is  high:  the losses are \emph{intermittent}.
The tail (\ref{tail})  is proportional to $a$ since it is  dominated   by losses in any one link along the path.  On the other hand, this link is shared by a large number of paths as illustrated in Fig.~\ref{Fig1}.

The tail  (\ref{tail})  is irrelevant for $  \left\langle \Phi \right\rangle_\mathrm{c}$ if $\delta>0$ but dominates higher moments of intermittent losses if $\delta<\frac{1}{2} $. Remarkably, in the SF models of the Internet as well as in  direct measurements of its link load \cite{Goh:01,Goh:02,Vazquez:02,Boccaletti:06,Dorogovtsev:08,Dimitropoulos:05} the exponent $2+\delta$ in Eq.~(\ref{PB}) lies between $ 2.0$ and $2.3$,  {i.e.} obeying $0<\delta<\frac{1}{2} $. The relative values of all  higher moments in this mode are large: e.g., $\sqrt{\langle{\Phi ^2}\rangle_\mathrm{c}}/\langle{\Phi }\rangle_\mathrm{c}\sim a^{1/2-\delta}(\varphi_0/a\gamma)^{1-\delta}  \gg1 $.

This relatively high magnitude of higher moments means  that losses occur in groups (intermittency).  Indeed, the \emph{fraction} of dropped \emph{pairs} can be represented using Eq.~(\ref{Phin}) as $ \frac2{K(K-1)}\sum_{k<j } \delta_k\delta_j$. It is approximately equal to $\Phi _i^2({T})$ if  a typical \emph{number} of lost packets, $\eta_iK_i$, is large. Similarly, the fraction of  $n$-tuples of dropped packets  is equal to $\Phi_i^n(T)$ (for $n\ll \eta_iK_i$). Then the \emph{small} probability of loosing, e.g., three packets scale, $\left\langle \Phi^3 \right\rangle_\mathrm{c}$, is much bigger  on the mesoscopic time than the probability of three independent loss events, $\left\langle \Phi \right\rangle_\mathrm{c}^3$.

The intermittency is the reason of a strong adverse effect of the fluctuations on network feedback. We illustrate this for the feedback mechanism provided by an idealized  transmission  control protocol (TCP) which handles most of data transfer. It works in cycles, each consisting of sending a group of  $W$ packets from the source and receiving acknowledgments of their delivery at the destination \cite{Kurose:10}. The transmission rate  equals $W/t_0$ where  the cycle duration $t_0$ is normally the round trip time.    On establishing a connection, $W$ linearly increases in time  until it reaches a steady-state value determined, in the absence of losses,  by the end-user resources. If a loss is detected during the cycle, the protocol  halves $W$. If losses occur in a few nearby cycles, $W$ is multiplicatively reduced. As it can increase only additively, such a multiple-loss event results in delays noticeable to the end user. Indeed, at the onset of congestion a typical round trip time is governed by a queue in a single full buffer \cite{Kurose:10} and is of order  $t_0=0.25$s. As in free-flow regime $W\gtrsim100$, the time of resuming normal rate of service could be tens of seconds.

Hence, it is crucial to know whether the protocol operation time scale at the onset of congestion,  $\varphi_0^2(t_0)\!\equiv\! \tau/t_0 $, falls into the mesoscopic mode where the relative probability of multiple losses is high. To this end note that the memory buffer size $c_i$ of any link is related to its capacity (maximal sending rate) $r_i$ by the engineering  `rule of thumb' \cite{Kurose:10}, $c_i=t_0 r_i$, ensuring any full buffer to empty during the same time $t_0$.  As $\tau_ir_i \lesssim1$ at the congestion threshold,  we find $\varphi_0^2(t_0) \gtrsim1/c$. We show this region of protocol operations as the hatched area in Fig.~\ref{Fig2} which spreads by many orders in magnitude over  $\gamma ^{-1}$.

The criticality width $\gamma $  is not directly measurable but is bounded. The upper bound in Fig.~\ref{Fig2}, $\gamma^{-1}\lesssim c$, is approaching the perfect design (full utilization), as explained after Eq.~(\ref{F}). The lower bound, $\gamma ^{-1}\gg aW$, is determined by the condition $\left\langle \Phi \right\rangle_\mathrm{c}\sim a\gamma\ll 1/W$: if it were not fulfilled, at least one packet per cycle would be lost resulting in a sharp reduction of the transmission rate after a few cycles, i.e.\ strong congestion.
Assuming $c\sim10^6$ for the typical buffer size (in  packets) \cite{Kurose:10}, and $a\sim10$ for the average number of links in an end-to-end path across  the Internet,  \cite{Albert:02,BBV:08} we see that $aW$ is close to $\sqrt{c} $, so that almost the entire hatched area corresponds to the mesoscopic mode of intermittent losses.

 On the macroscopic time scale (at the bottom of the hatched area) the probability $\left\langle \Phi^2 \right\rangle_\mathrm{c}$ of detecting two loss indicators in two consecutive cycles is simply equal to $\left\langle \Phi \right\rangle_\mathrm{c}^2$. However, for larger $\gamma^{-1}$ the ratio $ {\langle{\Phi ^2}\rangle_\mathrm{c}}/\langle{\Phi }\rangle_\mathrm{c}^2 $ increases reaching  $ c^{1-\delta}/a$
at the top, where $\gamma \sim c^{-1} $. There it  varies from $10^{2}$ for $\delta=\frac{1}{2} $ to $10^{5}$ for $\delta\to0$. Such a ratio for detecting three loss indicators in three consecutive cycles is even more striking, varying from $10^6$ to $10^9$. Conversely,  the same multiple-loss indicators  may correspond to very different \emph{average} losses  $a\gamma $. And it is the \emph{time-independent} average loss which matters since the intermittent fluctuations would die out with time as operations move from the dangerous dark-shaded (red online) area to the safe light-shaded (green online) one, see Fig.~\ref{Fig2}. Hence, it is of great importance to design protocols capable of  avoiding the overestimation of nascent losses by identifying in which loss mode the network operates and adjusting accordingly. To this end one needs to distinguish between single and multiple packet loss events within one cycle -- the information which is not collected in normal TCP operations.

The key features of our model might be characteristic of complex networks other than the Internet. First, if link (or node) operations in a network can be described by a finite-capacity model, it will suffer from  local congestion fluctuations at the threshold of capacity  due to inevitable input noise. Secondly, if such a network is heterogeneous, with a PL load distribution, the fluctuations would be greatly enhanced  on highly-loaded network elements.
Finally, the fluctuations become really dangerous  when they are misinterpreted by a network feedback mechanism -- the transmission control protocol in the case of the Internet. This mechanism is  specific for different types of network and whether it may trigger fluctuation-induced congestion requires \emph{ad hoc} considerations.

This work was supported by the EPSRC Grant No.\
EP/E049095/1.

\end{document}